\documentclass{aa}
\usepackage[varg]{txfonts}
\usepackage{graphicx}
\usepackage{natbib}
\usepackage{lipsum}
\bibpunct{(}{)}{;}{a}{}{,} 

\begin{document}
\title{Long-term influence of asteroids on planet longitudes and chaotic
dynamics of the solar system}

\titlerunning{Long-term chaotic influence of asteroids on planet longitudes}

\author{F. Bouchet\thanks{freddy.bouchet@ens-lyon.fr}
\and E. Woillez\thanks{eric.woillez@ens-lyon.fr}}

\institute{Univ Lyon, Ens de Lyon, Univ Claude Bernard, CNRS, Laboratoire de Physique, F-69342 Lyon, France}

\date{Received date /
Accepted date }

\abstract{The aim of this paper is to compare different sources of stochasticity
in the solar system. More precisely we study the importance of the
long term influence of asteroids on the chaotic dynamics of the solar
system. We show that the effects of asteroids on planets is similar
to a white noise process, when those effects are considered on a time
scale much larger than the correlation time $\tau_{\varphi}\simeq10^{4}$ yr of asteroid trajectories.
We compute the time scale $\tau_{e}$ after which the effects of the
stochastic evolution of the asteroids lead to a loss of information
for the initial conditions of the perturbed Laplace\textendash Lagrange
secular dynamics. The order of magnitude of this time scale is precisely
determined by theoretical argument. This time scale should be compared
with the Lyapunov time $\tau_{i}$ of the solar system without asteroids
(intrinsic chaos). We conclude that $\tau_{i}\simeq10\,\text{Myr}\ll\tau_{e}\simeq10^{4}\,\text{Myr}$,
showing that the external sources of chaoticity arise as a small perturbation
in the stochastic secular behavior of the solar system, rather due to intrinsic
chaos.} 

\keywords{Celestial mechanics -- Minor planets, asteroids: general}

\maketitle

\section{Introduction\label{sec:Introduction}}

The numerical integration of the secular equations of motion for the
eight planets of the solar system including the moon \citep{Laskar1989,Sussman_Wisdom_1992_chaotic}
has shown that the solar system is chaotic and its Lyapunov time $\tau_{i}$
has been estimated around 10 Myr. The chaos is significant enough
such that a single integration of the equations of motion is not at
all representative of the state of the solar system after 10 Myr.
One property of chaotic motion is that it increases exponentially
any difference in the initial positions of the planets. Therefore,
an error of few meters in the initial positions of the planets leads
on tens of Myr time scale to a complete indetermination on the actual
position of the planets. This has been confirmed by more and more accurate numerical integrations of the solar system dynamics without averaging and including perturbations induced by the largest asteroids \citep{laskar2011la2010,fienga2009inpop08}.

It is a natural question to address the effect of external sources
of chaoticity on the dynamics of the planets. Those external sources
could be e.g. the comets, the asteroids, or even the radiative pressure
from the sun. From all those external sources, a computation of orders
of magnitude shows that the asteroids should be the main ones. The
asteroid belt is a very good candidate for white noise on the planetary
dynamics, because asteroids, especially the larger ones Ceres, Pallas
and Vesta, are large enough to exert a non negligible gravitational
interaction on the planets. Moreover numerical studies by \cite{laskar_Gastineau_2011_AA_strong}
have shown that their dynamics is chaotic with a Lyapunov time $\tau_{\varphi}\simeq 10^4$ yr much
smaller than the Lyapunov time of 10 Myr of the solar system. Numerical
simulations of the dynamics of the solar system including the asteroids,
compared to others without asteroids \citep{laskar_Gastineau_2011_AA_strong},
show that they can indeed change the secular behavior of planetary
orbits, however it has been observed that simulations with asteroids
do not affect significantly the Lyapunov time of the solar system \citep{laskar_Gastineau_2011_AA_strong}. 

Nevertheless, a systematic investigation of the influence of asteroids
on planets has not been done yet. We have numerical evidences that asteroids perturbs planetary motion on tens of Myr time scale, but we don't know which physical parameters do control this time scale.
The aim of this paper is to answer this question, and to give a theoretical support to numerical simulations. We find that the
dominant effect of asteroids is on planetary mean longitudes. The first result
is that, if one would neglect secular evolutions of the orbital parameters,
the error on the longitude of Mars would be described by a superdiffusion,
with standard deviation growing like $\left(t/\tau_{\text{diff}}\right)^{3/2}$
, with
\begin{equation}
\tau_{\text{diff}}\propto\left(\left(\frac{m_{a}}{M_{S}}\right)^{2}\frac{\mathcal{G}M_{S}}{a_p^{3}\tau_{\varphi}}\right)^{-1/3},\label{eq:tau_diff_dimension}
\end{equation}
 where $\mathcal{G}$ is Newton's constant of gravitation, $m_{a,}\,M_{S}$
are the masses of the asteroid and the sun respectively, $a_p$ is the
semi-major axis of Mars, and $\tau_{\varphi}$ is the Lyapunov time
of the asteroid. Altogether, the magnitude of $\tau_{diff}$ is of
order of $20$ Myr. The second result is that the asteroids influence
on the secular dynamics leads to a superdiffusion of the orbital parameters eccentricities and inclinations,
with standard deviation growing like $\left(t/\tau_{\text{sec}}\right)^{3/2}$
, with
\begin{equation}
\tau_{\text{sec}}\propto\left(\left(\frac{m_{a}}{M_{S}}\right)^{2}\frac{\nu_{sec}^{2}}{\tau_{\varphi}}\right)^{-1/3},\label{eq:tau_sec_dimension}
\end{equation}
where $\nu_{sec}$ is the typical order of magnitude of the  secular frequencies of the Laplace\textendash Lagrange
system, about a few arcsec/yr. We can then estimate that $\tau_{sec}$
is of order of $10$ Gyr. \\

At a conceptual level, it may be useful to discuss the meaning of
stochasticity for deterministic dynamical systems. First, for times
much longer than the inverse of the largest Lyapunov exponent of a
dynamical system (what we call the Lyapunov time), a chaotic dynamics
is hardly distinguishable from a stochastic dynamics. One can consider
that two deterministic trajectories in the dynamical system differing
in their initial conditions are like two independent realizations
of a stochastic process. Precise mathematical results can be obtained
in several frameworks. For instance let us consider the case of two
coupled deterministic systems evolving on distinct fast and slow time scales, assuming
that the fast deterministic system is chaotic. One can prove mathematically
that in the limit of a large time scale separation, the effect of
the fast deterministic system is equivalent to the effect of a white
noise which properties are related to the statistics of the fast chaotic
dynamics. Unveiling the required hypothesis for such results to be
valid is a difficult and fascinating part of ergodic theory (see for
instance \citep{kifer2004averaging,kifer2009large} and reference therein).
The key message, which has been commonly accepted in statistical mechanics
for several decades, is that under generic conditions, there is no
fundamental differences between deterministic complex chaotic dynamics
and stochastic ones, beyond the very difficult and subtle mathematical
problems (see for instance \citep{GallavotiScholar,gallavotti1995dynamical,ruelle1980measures}).
As an illustration of this idea, \cite{pavliotis2008multiscale} have
shown that a pendulum coupled to the chaotic dynamics of the Lorentz system
is equivalent to a stochastic process. Another classical example is
the relation between Brownian motion and the deterministic dynamics
of atoms: a big particle coupled to a large number of smaller particles
has a stochastic dynamics which is very accurately described by the
Langevin equation for Brownian motion. In this latter example, the
small particles act as a white noise force on the large particle. 

The case of the coupling between a fast stochastic dynamics and a
slow dynamics is less technical from a mathematical point of view
than the deterministic one, while equivalent at a formal level. The
related tools, known as stochastic averaging are described in classical
textbooks \citep{pavliotis2008multiscale,gardiner1985stochastic,freidlin1984random}.
When a chaotic dynamics is what we call ``mixing'', which means
that the system looses fast enough the memory of its initial condition,
it is statistically equivalent to a white noise when it is considered 
on a time scale much longer than the mixing time. The properties
of the white noise are given by a Green\textendash Kubo formula, as we explain
in Sect. \ref{sec:Diffusion-over-the} . This result will be the
main theoretical starting point of our analysis. 

The importance of stochastic approaches for the solar system dynamics
has been understood a very long time ago. For instance \cite{laskar2008chaotic}
integrated numerically 1000 trajectories of planets of the solar system
with small differences in the initial conditions in order to compute
numerically the evolution of the probability distribution functions
of the eccentricities and inclinations of the main planets. More recent
work include \citep{tremaine2015statistical,mogavero2017addressing}.
Mogavero tried to reproduce the probability
distributions of planets numerically computed by \cite{laskar2008chaotic} using simply
the hypothesis of equiprobability in phase space and taking into account
the integrals of motions. The work of \cite{batygin2015chaotic} about
the chaotic motion of Mercury is, to our knowledge, the first one
in the context of celestial mechanics that models a chaotic Hamiltonian
dynamics with a stochastic process on a long time scale, where the
order of magnitude of the stochastic process properties are discussed
precisely. The case of the stochastic effect of the asteroids on the
planets has motivated the theoretical work of \cite{lam2014stochastic}. They assumed that the
effect of asteroids on planetary Keplerian motions is equivalent
to that of a white noise acting on an integrable system. In the present
paper, we show that this assumption is satisfied when the time scale
considered is much larger than the Lyapunov time $\tau_{\varphi}$
of asteroids, and we give the theoretical expression and the order
of magnitude of the white noise process. If the white noise has an
amplitude $\delta$, Thu Lam and Kurchan have shown
that the Lyapunov time of the perturbed system was scaling like $\left(\frac{1}{\delta}\right)^{1/3}$.
The first technical part of our paper, is very similar to the one
of \cite{lam2014stochastic}. Rather than studying the Lyapunov exponent,
we study the diffusion of planetary mean longitudes, which is more relevant if one
is interested in the evolution of the probability distribution functions.
The relevant time scales are however similar: the Lyapunov time (the
inverse of the Lyapunov exponent) is of the same order of magnitude
as the diffusion time $\tau_{diff}$, of order of $20$ Myr (see table
(\ref{tab:asteroids})). The second technical part of our paper connects
correctly for the first time the deterministic dynamics of the asteroids
with the model with white noises, using stochastic averaging. This
allows to compute correctly for the first time the order of magnitude of $\delta$, and to actually
compute the relevant time scales for the solar system. We apply these
results to both the diffusion of the planetary mean longitudes and the orbital elements for the Laplace\textendash Lagrange dynamics.\\

We give in Sect. \ref{sec:The-theoretical-framework} the precise
mathematical formulation of the question we are interested in. Sect.
\ref{sec:The-theoretical-framework} may thus appear a bit abstract
from the point of view of celestial mechanics, but it is essential
to understand the scope of the present work, which is larger, as we
explain in this introduction, than answering only the question of
long-term influence of asteroids and could be applied to many dynamical
systems. In Sect. \ref{sec:A-simplified-model} we introduce a simplified
Hamiltonian model to describe the dynamics of Mars under the influence
of asteroids, and we justify why this model is relevant to compute
orders of magnitude. The stochastic method is explained in Sect.
\ref{sec:Diffusion-over-the}. This section is a bit technical although
most of the difficulties are skipped and done in appendix. Sect.
\ref{sec:Orders-of-magnitude} contains the main results of the paper,
the computation of the time scale $\tau_{diff}$. We then investigate
the influence of asteroids on planetary secular motion in Sect.
\ref{sec:Influence-on-secular} and we conclude our discussion in
Sect. \ref{sec:Conclusion}. In order to help the reader, we have gathered the main notations appearing in our article in appendix \ref{sec:Notations}.

\section{The theoretical framework to compare intrinsic and extrinsic chaos\label{sec:The-theoretical-framework}}

In this section, we formulate the problem in a more precise mathematical
set-up. The present discussion goes far beyond the particular problem
of the influence of asteroids on planetary dynamics. The influence
of asteroids is one simple application of our results, but the mathematical
framework should find many other applications in celestial mechanics.

We are considering a dynamical system of the form
\begin{equation}
\frac{{\rm d}x}{{\rm d}t}=f_{0}(x)+\eta f_{1}(x)+\epsilon g(x,t).\label{eq:complete framework}
\end{equation}
The system has two small parameters $\eta\ll1$ and $\epsilon\ll1$,
$x$ is a multidimensional vector. The field $f_{0}+\eta f_{1}$ depends
only on $x$, it can be considered as the \emph{intrinsic dynamics
}of $x$. The field $g$ depends both on $x$ and the time, it acts
as an \emph{external perturbation} on the system.

We assume that the zeroth order dynamics defined by 
\[
\frac{{\rm d}x}{{\rm d}t}=f_{0}(x)
\]
is an integrable Hamiltonian dynamics. When the small perturbation
$\eta f_{1}$ is added, we assume that the intrinsic dynamics
\begin{equation}
\frac{{\rm d}x}{{\rm d}t}=f_{0}(x)+\eta f_{1}(x)\label{eq:intrinsic dynamics}
\end{equation}
 is chaotic with a Lyapunov time $\tau_{i}$ (the Lyapunov time is
defined as the inverse of the largest Lyapunov exponent). In the present
paper, we will consider two cases for the dynamics (\ref{eq:intrinsic dynamics}).
Sect. \ref{sec:A-simplified-model} will present the case where
the integrable dynamics $f_{0}$ is the Keplerian dynamics of planets,
$\eta f_{1}$ being the perturbative function coming from the gravitational
interactions between the planets. In Sect. \ref{sec:Influence-on-secular},
$f_{0}$ corresponds to the secular system of Laplace\textendash Lagrange
and $\eta f_{1}$ gathers all non-linear terms of order two and more
in eccentricities and inclinations. From the results of the work of
\cite{Laskar1989}, we know that the full non-linear secular dynamics
is chaotic with an intrinsic Lyapunov time $\tau_{i}$ of the order
of 10 Myr. 

External sources may add a small perturbation on the intrinsic dynamics (\ref{eq:intrinsic dynamics}) described by the term $\epsilon g(x,t)$.
In the present paper for example, we are considering the asteroids
as an external source of noise for planets of the solar system. The
intrinsic dynamics is already chaotic, and therefore the complete
system (\ref{eq:complete framework}) is also chaotic. On the other
hand, the integrable system perturbed by the external source of noise
\begin{equation}
\frac{{\rm d}x}{{\rm d}t}=f_{0}(x)+\epsilon g(x,t)\label{eq:extrinsic dynamics}
\end{equation}
is chaotic with a Lyapunov time $\tau_{e}$. In the case (\ref{eq:extrinsic dynamics}),
the chaos is due to the external perturbation. 

As we explained in the introduction, if the perturbation $g(x,t)$
satisfies a \emph{mixing }condition, which means that its time correlation
function for any fixed value of $x$ decreases fast enough, the deterministic
system of equations (\ref{eq:extrinsic dynamics}) is equivalent to
a Langevin process
\begin{equation}
\frac{{\rm d}x}{{\rm d}t}=f_{0}(x)+\xi(x,t),\label{eq:stochastic reduction}
\end{equation}
where $\xi(x,t)$ is a white noise, which amplitude can be expressed
with the properties of $g(x,t)$ with a Green-Kubo formula. This point
will be discussed in Sect. \ref{sec:Diffusion-over-the}. Once the
system (\ref{eq:extrinsic dynamics}) has been written as a stochastic
process (\ref{eq:stochastic reduction}), it is quite easy to give the order of magnitude of $\tau_{e}$.
Equations (\ref{eq:intrinsic dynamics}-\ref{eq:extrinsic dynamics})
define two different regimes depending on the Lyapunov times $\tau_{i}$
and $\tau_{e}$ :
\begin{enumerate}
\item The regime $\tau_{i}\ll\tau_{e}$ defines a regime of intrinsic chaos.
On a time of order of the internal Lyapunov time $\tau_{i}$, the
effect of the external perturbation $\epsilon g$ is small. Thus,
the probability distributions of the variable $x$ are essentially
the same, to leading order in $\epsilon$, for the full system (\ref{eq:complete framework})
and for the intrinsic dynamics (\ref{eq:intrinsic dynamics}).
\item If on the contrary $\tau_{e}\ll\tau_{i}$, then the external perturbation
creates chaos in the integrable system, in the sense that the system
looses the memory of its initial condition before the intrinsic chaos
can develop. For intermediate times between $\tau_{e}$ and $\tau_{i}$,
the complete dynamics (\ref{eq:complete framework}) can thus be described
by a stochastic process, and the probability distributions may be
strongly influenced by the external perturbation.
\end{enumerate}
Using the present framework, our question formulates in a very simple
way. Let the complete set of equations (\ref{eq:complete framework})
be the equations for planetary motion of the solar system perturbed
by the asteroid belt, are we in the regime of \emph{dominant intrinsic
chaos} with $\tau_{i}\ll\tau_{e}$ or in the regime of \emph{external
source of chaos }with $\tau_{e}\ll\tau_{i}$? What is then the order
of magnitude of $\tau_{e}$ describing the interactions between asteroids and planets?

\section{A simplified model for planet-asteroid interactions\label{sec:A-simplified-model}}

For times smaller than the Lyapunov time of the solar system, the
secular motion of planetary orbits is very accurately approximated by the quasiperiodic
solution of the Laplace\textendash Lagrange equations. Except for the smallest
planet Mercury, the planetary orbital elements eccentricities and inclinations remain very small (less than 0.15 for the eccentricity,
and less than 10 degrees for the inclination) in the Myr time scale
\citep{laskar2008chaotic}. The computations to be performed in the following could be
done without fondamental difficulties for elliptic and inclined trajectories.
However solving these equations for the elliptic motion is technically much
more tedious than for restricted planar and circular motions, and
it would not change the orders of magnitude to leading order in eccentricities
and inclinations. To study the order of magnitude of the perturbation
induced by the asteroids on the planets, we will thus introduce a
simplified model where the orbits of celestial bodies are circular
and coplanar. To describe the  motion of the planet, we thus keep only the
two orbital elements mean longitude $\lambda$ and semi-major axis $a$.
\begin{table*}

\caption{Physical properties of the main asteroids of the belt, Ceres, Vesta
and Pallas. $M_{S}$ is the mass of the sun. The Lyapunov times are
taken from \cite{laskar_Gastineau_2011_AA_strong}}

\label{tab:Physical-properties-of}

\centering
\begin{tabular}{c c c c c c}
\hline 
 & non-dimensional mass $\epsilon=\frac{m}{M_{S}}$ & semi-major axis (u.a) & eccentricity & inclination & Lyapunov time $\tau_{\varphi}$ (yr)\\
\hline 
\hline 
Ceres & $4.7*10^{-10}$ & $2.8$ & 0.076 & $11\text{\r{ }}$ & 28900\\

Vesta & $1.35*10^{-10}$ & $2.4$ & $0.09$ & $7\text{\r{ } }$ & 14282\\

Pallas & $1.05*10^{-10}$ & $2.7$ & 0.23 & $35\text{\r{ }}$ & 6283\\
\hline 
\end{tabular}

\end{table*}

The main perturbation will be on planet Mars whose orbit is the
closest to the asteroid belt, but our method can be applied straightforwardly
to other planets, using the correct orbital elements. Most of the
mass of the asteroids comes only from the contribution of the largest
ones Ceres, Vesta and Pallas (about 58 $\%$ of the total mass of the asteroids). The physical properties of those three
asteroids is summarized in table \ref{tab:Physical-properties-of}.
In our simplified model, we will thus retain only the planet Mars
and one asteroid. Asteroids have chaotic motions because of gravitational perturbation by the planets, but also because of interactions between each other as shown by \cite{laskar_Gastineau_2011_AA_strong}.They are thus not independent. Yet,
their motion is decorrelated after the Lyapunov time, that's
why we can do the hypothesis that the perturbation of asteroids are
independent, and we will thus add the individual contributions of
each asteroid in the final result to obtain the right order of magnitude.
The simplified model of the planet Mars perturbed by one asteroid
is described by the Hamiltonian
\[
H=-\frac{\mathcal{G}M_{S}m_{p}}{2a_{p}}-\frac{\mathcal{G}M_{S}m}{2a}-\frac{\mathcal{G}mm_{p}}{\left|a_{p}e^{i\lambda}-ae^{i\varphi}\right|},
\]
where $M_{S},m_{p},m$ stand for the masses of the Sun, Mars, and
the asteroid respectively, $\lambda,\varphi$ are the mean longitudes of
Mars and the asteroid, and $a_{p}$ and $a$ their respective semi-major axis.
The Sun is considered as fixed, and we take the real mass $m$ instead
of the reduced mass $\frac{1}{\beta}=\frac{1}{m}+\frac{1}{M_{S}}$
as should be done rigorously. 

In physical units, it is difficult to see where is the small parameter
in the problem. Therefore, we rescale all physical variables. The
mass of the sun is taken as the unit mass, $M_{S}=1$, and the reduced
mass $\epsilon:=\frac{m}{M_{S}}$ of the asteroid is thus very small
(table \ref{tab:Physical-properties-of}). Then we change the units
for time and length, the astronomic unit is the new length scale,
and we choose the units of time such that the actual Keplerian period
of Mars is $2\pi$. From the relation $\frac{T_p^{2}}{a_{p}^{3}}=\frac{4\pi^{2}}{\mathcal{G}M_{S}}$
we get the new time unit $\frac{a_{p}^{3}}{\mathcal{G}M_{S}}=1.$ Denoting $n_p:=\frac{2\pi}{T_p}$ the Keplerian pulsation of Mars, this means that $n_p=1$.
Finally, let $\Lambda:=\epsilon_{p}\sqrt{a_{p}}$ be the canonical
momentum associated to $\lambda$, we do the canonical change of variables
$H\leftarrow\frac{H}{\epsilon_{p}},\Lambda\leftarrow\frac{\Lambda}{\epsilon_{p}}$.

In the work of \cite{laskar_Gastineau_2011_AA_strong}, the orbits
of asteroids have been shown to be chaotic and the Lyapunov time $\tau_{\varphi}$ on
their longitudes has been computed numerically. The Lyapunov times
for the three main asteroids are given in table \ref{tab:Physical-properties-of}.
In the present model of planet Mars perturbed by one asteroid, it is the chaos of the asteroid's motion
that breaks the periodic Keplerian regular motion of Mars. Of
course the chaotic motion of the asteroid does not come from the influence
of Mars alone, but rather from interaction and close encounters with other asteroids
and on the influence of the giant planets. The retroactive influence
of Mars on the asteroid can be considered as negligible compared to
the influence of Jupiter or Saturn, and the interaction with Mars
cannot substantially change the characteristics of the orbit of the
asteroid, in particular its Lyapunov time. That's why we do not solve the equations of motion for the asteroid.
To capture the physical
phenomena coming from the gravitational interaction between Mars and
the asteroid, the trajectory of the asteroid has to be considered
as an input function in the model which properties come from more
precise numerical studies. We thus take the semi-major axis $a$ as
a constant and the phase of the asteroid $\varphi(t)$ as a function
of time with correlation time of the order of the Lyapunov time $\tau_{\varphi}$.
The trajectory of Mars in our model is thus a functional of the trajectory $\varphi (t)$ of the asteroid.
The reader should always bear in mind that those strong hypothesis
are done to the aim of giving orders of magnitude and not precise
quantitative results.

The Hamiltonian of the simplified model we will study writes
\begin{equation}
H(\Lambda,\lambda,t):=-\frac{1}{2\Lambda^{2}}-\frac{\epsilon}{\left|\Lambda^{2}e^{i\lambda}-ae^{i\varphi(t)}\right|}.\label{eq:simple Hamiltonian}
\end{equation}

The Hamiltonian (\ref{eq:simple Hamiltonian}) does not conserve the
total energy and angular momentum of the asteroid and Mars. But to
first order in $\epsilon$, energy and angular momentum are the ones
given by the Keplerian orbit and depend thus only on the canonical
momentum $\Lambda$. Our model is consistent if the change in energy
and angular momentum occurs on a time scale much larger than the time
we are considering for the perturbation of Mars. This point will
be checked a posteriori in Sect. \ref{sec:Orders-of-magnitude}
when we will obtain the time $\tau_{diff}$ over which the perturbation
on Mars becomes large.

In the final paragraph of this section, we discuss more precisely in which sense the position
of Mars should be considered as a stochastic variable associated to
a probability distribution. What is usually done in numerical simulations
of chaotic planetary motion is to choose a large number of initial
conditions differing only by a small shift in the initial position.
Because of chaos, the different trajectories do not stay close together
but separate exponentially fast on a time scale given by the Lyapunov
exponent. After a time sufficiently long compared to the Lyapunov
time, the positions of all trajectories give a distribution. This
distribution is an estimation of all possible positions that could
be reached from a uniform distribution on a very small set of initial
conditions. In this sense, it is a probability distribution. In the
simplified model (\ref{eq:simple Hamiltonian}), we do not fix the
initial position of the asteroid. We will thus study the motion of
Mars for different possible realizations of the function $\varphi(t)$
and we take for $\varphi(0)$ a uniform distribution over the range
$[0,2\pi]$. This choice is done because the incertitude on the longitude
of the asteroid becomes complete after a time large enough compared to the Lyapunov time $\tau_{\varphi}$.
 The position of Mars has a probability distribution because
it is conditioned by the realizations of the stochastic function $\varphi(t)$.
We really emphasize that the trajectories of Mars will not separate
exponentially with time, as would have been the case if we had just
consider a set of very close initial conditions for the position of
Mars and one single function $\varphi(t)$. In our model, $\varphi(t)$
is a random function with a probability distribution $\mathbb{P}[\varphi]$.
Given this probability distribution, we want to obtain the probability
distribution of the canonical variables $\Lambda,\lambda$. Instead
of an exponential separation, we will get a diffusive behavior as
will be shown in the next section.

\section{The stochastic longitude evolution as a diffusion\label{sec:Diffusion-over-the}}

From the Hamiltonian (\ref{eq:simple Hamiltonian}), we get the set
of Hamilton equations for $\lambda,\Lambda$ as 
\begin{eqnarray}
\frac{{\rm d}\lambda}{{\rm d}t} & = & n_p(\Lambda)+\epsilon\frac{\partial G}{\partial\Lambda}(\Lambda,\lambda-\varphi(t))\label{eq:model}\\
\frac{{\rm d}\Lambda}{{\rm d}t} & = & -\epsilon\frac{\partial G}{\partial\lambda}(\Lambda,\lambda-\varphi(t)),\nonumber 
\end{eqnarray}
where we have introduced the Keplerian pulsation $n_p(\Lambda):=\frac{1}{\Lambda^{3}}$
and the gravitational interaction $G(\Lambda,\lambda-\varphi):=\frac{-1}{\left|\Lambda^{2}e^{i(\lambda-\varphi)}-a\right|}$
. For technical reasons, it is convenient to consider the variable
$p:=n_p(\Lambda)$ instead of $\Lambda$ and write the Hamilton equations
(\ref{eq:model}) as
\begin{eqnarray}
\dot{\lambda} & = & p+\epsilon\frac{\partial G}{\partial\Lambda}(\Lambda,\lambda-\varphi(t))\label{eq:model-bis}\\
\dot{p} & = & -\epsilon\frac{\partial n_p}{\partial\Lambda}(\Lambda)\frac{\partial G}{\partial\lambda}(\Lambda,\lambda-\varphi(t)).\nonumber 
\end{eqnarray}
It is physically clear that to zeroth order in $\epsilon$, the motion
of $(p,\lambda)$ is simply a linear flow, with $p=p(0)$ and $\lambda(t)=\lambda(0)+p(0)t$.
Equations are then invariant with the change of variables $p\leftarrow p-p(0)$,
$\lambda\leftarrow\lambda-p(0)t$, provided the function $\varphi(t)$
is also changed as $\varphi(t)\leftarrow\varphi(t)-p(0)t$. This change
of variables means that we integrate out the Keplerian motion of Mars
and the new function $\varphi(t)$ represents the difference between
the mean longitude of the asteroid and the Keplerian mean longitude of Mars.

Now comes the important technical step in the calculation: from equations
(\ref{eq:model-bis}), it is not obvious on which time scale the integrable
motion of Mars will be perturbed. It should scale with $\epsilon$
but we still do not know the precise scaling at this step of the calculation.
Following the method proposed in \cite{lam2014stochastic}, we thus
introduce \emph{a priori} an exponent $\alpha>0$ and rescale the
time according to $t'=\epsilon^{\alpha}t$. The variables $\lambda$
and $p$ are also rescaled according to $\lambda'(t)=\lambda\left(\frac{t}{\epsilon^{\alpha}}\right)$
and $p'(t)=\frac{1}{\epsilon^{\alpha}}p\left(\frac{t}{\epsilon^{a}}\right)$
and $\Lambda'(t)=\Lambda\left(\frac{t}{\epsilon^{\alpha}}\right)$.
The equations for the rescaled variables are 
\begin{eqnarray}
\dot{\lambda}' & = & p'+\epsilon^{1-\alpha}\frac{\partial G}{\partial\Lambda}\left(\Lambda',\lambda'-\varphi\left(\frac{t}{\epsilon^{\alpha}}\right)\right)\label{eq:fast oscillation}\\
\dot{p}' & = & -\epsilon^{1-2\alpha}\frac{\partial n_p}{\partial\Lambda}(\Lambda')\frac{\partial G}{\partial\lambda}\left(\Lambda',\lambda'-\varphi\left(\frac{t}{\epsilon^{\alpha}}\right)\right).\nonumber 
\end{eqnarray}
To go one step further, we have to find the asymptotic behavior of
the two functions $\frac{\partial G}{\partial\lambda}\left(\Lambda',\lambda'-\varphi\left(\frac{t}{\epsilon^{\alpha}}\right)\right)$
and $\frac{\partial G}{\partial\Lambda}\left(\Lambda',\lambda'-\varphi\left(\frac{t}{\epsilon^{\alpha}}\right)\right)$
in the limit $\epsilon\rightarrow0.$ This limit is not at all trivial.
The averaging principle, which is classically used in celestial mechanics
to obtain the secular equations, states that the oscillating function
$\frac{\partial G}{\partial\lambda}\left(\Lambda',\lambda'-\varphi\left(\frac{t}{\epsilon^{\alpha}}\right)\right)$
should be averaged over the distribution of the fast angle $\varphi$.
However, it happens that the function $\frac{\partial G}{\partial\lambda}$
is periodic in $\varphi$, and its average over a uniform distribution
of $\varphi$ is zero. The averaging principle only tells us that
the semi-major axis is invariant to main order in $\epsilon$. The
semi-major axis is an adiabatic invariant because it
is a conserved quantity of the Keplerian fast motion, this result
is known in celestial mechanics as the theorem of Laplace\textendash Lagrange.
To obtain a non trivial variation of the semi-major axis, we have
to go to next order, beyond the averaging principle. 

For general fast oscillating functions $\frac{\partial G}{\partial\lambda}$,
there is no asymptotic expansion beyond the averaging principle. If
however the function $\frac{\partial G}{\partial\lambda}$ has a sufficiently
small correlation time, the limit exists and is a stochastic process.
The principle used to find the limit of $\frac{\partial G}{\partial\lambda}$
when $\epsilon$ goes to zero is called the homogenization process
\citep{gardiner1985stochastic,pavliotis2008multiscale}. The crucial
hypothesis is the one of decorrelation of the fast oscillating function.
In case of equations (\ref{eq:fast oscillation}), this hypothesis
is satisfied because of the presence of the asteroid mean longitude $\varphi(t)$.
The hypothesis on $\varphi$ is that it has a chaotic dynamics with
Lyapunov time $\tau_{\varphi}$. It is usually assumed that the Lyapunov
time is of the same order of magnitude as the correlation time of the chaotic trajectory.
If $\frac{1}{\epsilon^{\alpha}}\gg\tau_{\varphi}$, the theory of
stochastic averaging (\cite{gardiner1985stochastic} chapter 8) shows
that the function $\frac{\partial G}{\partial\lambda}\left(\Lambda',\lambda'-\varphi\left(\frac{t}{\epsilon^{\alpha}}\right)\right)$
is equivalent (it is said to be equivalent \emph{in law}) to a stochastic
process 
\begin{eqnarray}
\frac{\partial G}{\partial\lambda}\left(\Lambda',\lambda'-\varphi\left(\frac{t}{\epsilon^{\alpha}}\right)\right) & \underset{\epsilon\rightarrow0}{\sim} & \epsilon^{\alpha/2}\left[B(\Lambda')+\sqrt{D(\Lambda')}\xi(t)\right],\label{eq:white noise}
\end{eqnarray}
where $\xi(t)$ is the normal Gaussian white noise, $\left\langle \xi(t)\xi(t')\right\rangle =\delta(t-t')$,
and the coefficients $B$ and $D$ are given in terms of the correlation
function of $\frac{\partial G}{\partial\lambda}$ (see \cite{gardiner1985stochastic}
p189)
\begin{align}
B(\Lambda') & =\int_{0}^{+\infty}{\rm d}t\left\langle \frac{\partial^{2}G}{\partial\Lambda\partial\lambda}\left(\Lambda',\lambda'-\varphi\left(t\right)\right)\frac{\partial G}{\partial\lambda}\left(\Lambda',\lambda'-\varphi\left(0\right)\right)\right\rangle ,\label{eq:coefficients}\\
D(\Lambda') & =2\int_{0}^{+\infty}{\rm d}t\left\langle \frac{\partial G}{\partial\lambda}\left(\Lambda',\lambda'-\varphi\left(t\right)\right)\frac{\partial G}{\partial\lambda}\left(\Lambda',\lambda'-\varphi\left(0\right)\right)\right\rangle .\nonumber 
\end{align}
It should be noticed that the function $G$ depends on the variable
$\lambda-\varphi$, and as a result, the coefficients expressed in
(\ref{eq:coefficients}) do not depend on $\lambda'$ . The function
$\frac{\partial G}{\partial\Lambda}$ on the contrary, is not periodic
in $\varphi$ and a simple averaging principle is enough to give the
equivalent, $\frac{\partial G}{\partial\Lambda}\left(\Lambda',\lambda'-\varphi\left(\frac{t}{\epsilon^{\alpha}}\right)\right)\underset{\epsilon\rightarrow0}{\sim}\left\langle \frac{\partial G}{\partial\Lambda}\right\rangle (\Lambda)$,
where the average is done over $\varphi$.

Altogether, we can give a \emph{stochastic} equivalent of the system
of equations (\ref{eq:fast oscillation}) 
\begin{eqnarray}
\dot{\lambda}' & = & p'+\epsilon^{1-\alpha}\left\langle \frac{\partial G}{\partial\Lambda}\right\rangle (\Lambda')\label{eq:stochastic}\\
\dot{p}' & = & -\epsilon^{1-\frac{3}{2}\alpha}\frac{\partial n_p}{\partial\Lambda}(\Lambda')\left[B(\Lambda')+\sqrt{D(\Lambda')}.\xi(t)\right].\nonumber 
\end{eqnarray}
In the first equation of (\ref{eq:stochastic}), the perturbation
appears at order $\epsilon^{1-\alpha}$, whereas in the second equation
it is $\epsilon^{1-\frac{3}{2}\alpha}$. As $\alpha>0$, the term in the equation
for $p'$ is dominant. We choose the value $\alpha=\frac{2}{3}$,
and we can drop the term $\epsilon^{1-\alpha}\left\langle \frac{\partial G}{\partial\Lambda}\right\rangle (\Lambda)$
in the first equation. Forgetting the primes for the variables $\lambda'$
and $p'$ we finally write the stochastic equations for the dynamics
of Mars perturbed by an asteroid 
\begin{eqnarray}
\dot{\lambda} & = & p\label{eq:stochastic system}\\
\dot{p} & = & -\frac{\partial n_p}{\partial\Lambda}(\Lambda)B(\Lambda)+\frac{\partial n_p}{\partial\Lambda}(\Lambda)\sqrt{D(\Lambda)}\xi(t).\nonumber 
\end{eqnarray}

The last set of equations will describe to main order the diffusion
of $\lambda$ over a time scale $\frac{1}{\epsilon^{2/3}}$. This
result could seem counterintuitive: a glance at the initial equations
(\ref{eq:model}) could suggest that the perturbation should grow
as $\frac{1}{\epsilon}$, or at least as an entire power of $\frac{1}{\epsilon}$.
The strange exponent $\frac{2}{3}$ comes from the fact that the perturbation
over the semi-major axis $\Lambda^{2}$ is amplified on the mean longitude
$\lambda$ by Keplerian motion. The system (\ref{eq:stochastic system})
has an exact solution because $\Lambda$ should be considered as a
constant variable in the equation. In fact, the initial canonical
variable $\Lambda(t)$ evolves on a much longer time scale of order
$\frac{1}{\epsilon^{2}}$, and on the time scale $\frac{1}{\epsilon^{2/3}}$,
it has thus only very small variations. From the integration of (\ref{eq:stochastic system}),
we deduce that the probability distribution of $\lambda$ is a Gaussian
law. We can now compute its variance w.r.t the realizations of the
noise $\xi$. The term $-\frac{\partial n_p}{\partial\Lambda}(\Lambda)B(\Lambda)$
will give a deterministic contribution on $\lambda$ scaling like
$t^{2}$. The interesting part comes from the white noise, because
it makes the probability distribution of $\lambda$ spread over time.
We have $p(t)=-\frac{\partial n_p}{\partial\Lambda}(\Lambda)B(\Lambda)t+\frac{\partial n_p}{\partial\Lambda}(\Lambda)\sqrt{D(\Lambda)}W(t)$
where $W$ is the standard Brownian motion (also called the Wiener
process), and 
\begin{eqnarray*}
V_{\lambda}(t) & := & \boldsymbol{E}[\lambda^{2}(t)]-\boldsymbol{E}[\lambda(t)]^{2}\\
 & = & \left(\frac{\partial n_p}{\partial\Lambda}(\Lambda)\right)^{2}D(\Lambda)\int\int{\rm d}s{\rm d}s'\boldsymbol{E}[W(s)W(s')]\\
 & = & \left(\frac{\partial n_p}{\partial\Lambda}(\Lambda)\right)^{2}D(\Lambda)\int\int{\rm d}s{\rm d}s'\inf(s,s')\\
 & = & \frac{1}{3}\left(\frac{\partial n_p}{\partial\Lambda}(\Lambda)\right)^{2}D(\Lambda)t^{3}.
\end{eqnarray*}
This proves that $\lambda$ is \emph{``superdiffusive''}, because
the variance grows like $t^{3}$ instead of $t$ for standard Brownian
motion. Coming back to the original time, we can define a diffusive
time scale and write the mean variation $\Delta\lambda$ as 
\begin{align}
\tau_{diff} & =\left(\frac{1}{3}\epsilon^{2}\left(\frac{\partial n_p}{\partial\Lambda}\right)^{2}D(\Lambda)\right)^{-1/3},\label{eq:Diff_time}\\
\Delta\lambda(t) & =\left(\frac{t}{\tau_{diff}}\right)^{\frac{3}{2}}.\label{eq:delta-lambda}
\end{align}
A final remark will conclude this subsection: the asymptotic result
(\ref{eq:white noise}) shows that the interaction with chaotic asteroids
is equivalent to a white noise force acting on the planet, on a time
scale much larger than the Lyapunov time $\tau_{\varphi}$ of the
asteroid. (\ref{eq:coefficients}) give the properties of the noise,
and is the starting point to study the order of magnitude of the noise
amplitude, which is done in the next section. Our computation thus gives
a theoretical ground to the model studied in \cite{lam2014stochastic}
of an integrable dynamics perturbed by a white noise, and the correct order of magnitude for $B$ and $D$ in Eq. (\ref{eq:coefficients}) .

\paragraph{Comparison with Taylor-Aris dispersion}

\begin{figure}
\begin{centering}
\includegraphics[scale=0.55]{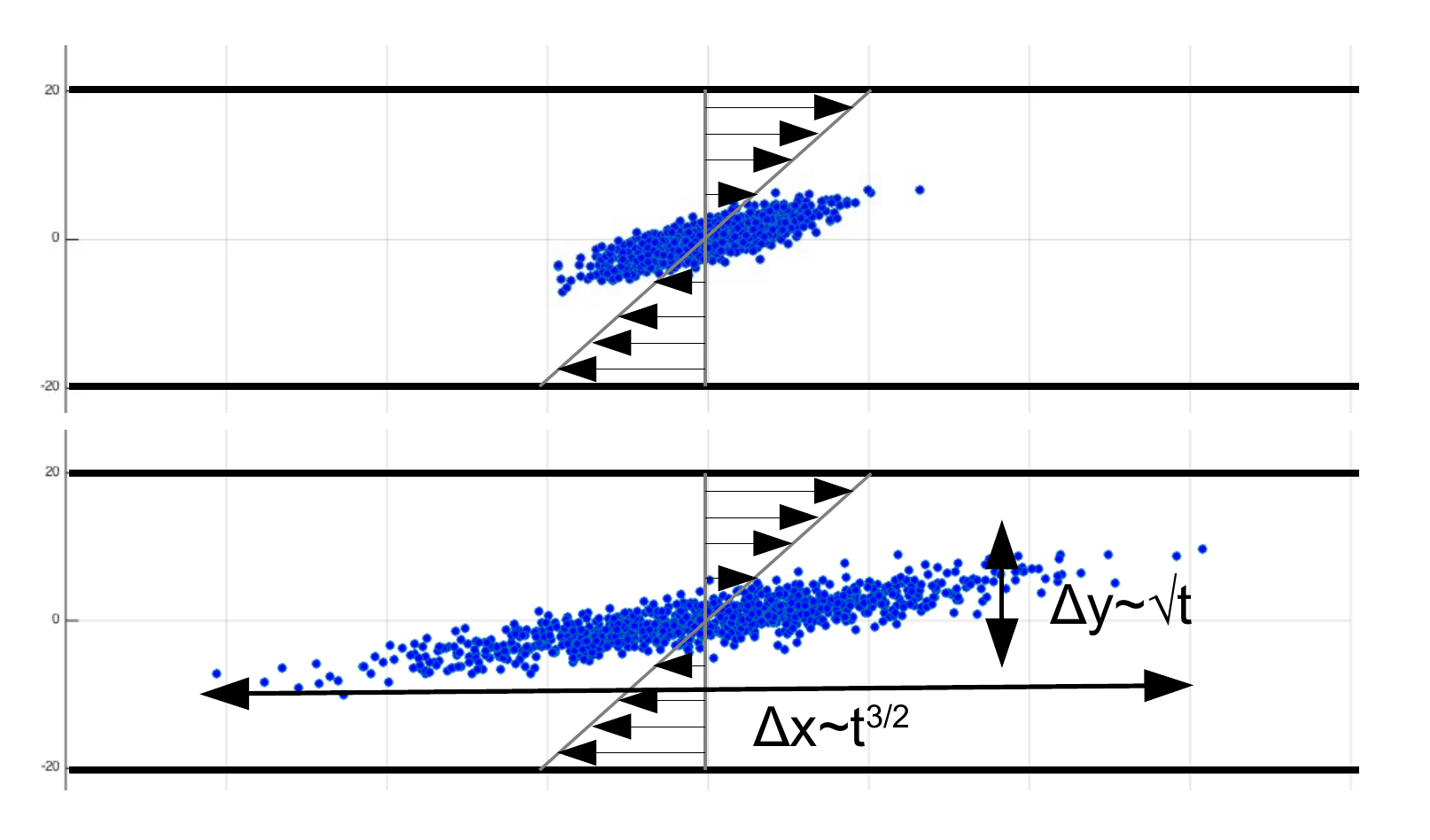}
\par\end{centering}
\caption{Dispersion of particles in a channel at $t=5$ (top) and $t=10$ (bottom).
The flow is indicated with arrows.\label{fig:dispersion of particles}}

\end{figure}

Although the results of this section could seem very technical at
first sight, the underlying physical mechanism is very simple. The
superdiffusion and the scaling $t^{\frac{3}{2}}$ of the dispersion
has been known for a long time in the hydrodynamics community where
it is called ``Taylor\textendash Aris '' dispersion. Let us describe
this phenomenon to illustrate the result (\ref{eq:Diff_time}-\ref{eq:delta-lambda}). Consider
a flow in a channel with linear velocity profile $v(y)=ny$ as displayed
on Fig. (\ref{fig:dispersion of particles}). At $t=0$, particles
are released in the middle of the channel at $y=0$. Particles will
diffuse along the x-direction and the y-direction. Along the y-direction,
we have a simple diffusion and $\left\langle y^{2}(t)\right\rangle \propto t$.
In the x-direction, the flow will amplify the diffusion: particles
going up at $y$ feel a velocity $ny$ and are carried forward in
the x-direction, whereas those going done at $-y$ feel the velocity
$-ny$ and are carried backward in the x-direction. Altogether, if
at time t particles have diffused over a range $\Delta y$, the dispersion
along $x$ will scale like $\Delta x=n\Delta yt$, and since $\Delta y\propto\sqrt{t}$
we find that $\Delta x\propto nt^{\frac{3}{2}}$. A diffusion scaling
with $t^{\frac{3}{2}}$ is an example of what we call a ``superdiffusion'', it
is illustrated on Fig.~(\ref{fig:dispersion of particles}). The
main result we prove in this paper is that the same mechanism happens
for planets. The gravitational interaction with the chaotic asteroid
causes a diffusion of the semi-major axis $a_p$. $a_p$
is the equivalent of the $y$ coordinate in Taylor-Aris dispersion.
The mean longitude $\lambda$ circulates at angular velocity - or Keplerian
pulsation- $n_p$. The Keplerian motion amplifies the diffusion
on $\lambda$ exactly as the flow on Fig.~(\ref{fig:dispersion of particles})
amplifies the dispersion on the $x$-axis. 

\section{Orders of magnitude for the diffusion coefficient, the diffusion and Lyapunov time scales\label{sec:Orders-of-magnitude}}

\paragraph{Scaling of the diffusion coefficient}

We want to evaluate the order of magnitude of the typical diffusion
time for the mean longitude of Mars, given by the formula (\ref{eq:Diff_time})
together with the theoretical expression of the diffusion coefficient
$D(\Lambda)$ in (\ref{eq:coefficients}). The difficult task comes
of course from the diffusion coefficient, because it involves the
correlation function of the derivative of $G$. The full computation
is reported in appendix \ref{sec:Computation-of-the}. The computation
depends only on an averaging over the phase of the asteroid. As the
motion of the asteroid is a chaotic function for which we do not have
an analytic expression, we have to do an hypothesis on how the phase
$\varphi$ of the asteroid differs from a simple Keplerian motion.
We have to take into account that the Lyapunov time of the phase is
given by $\tau_{\varphi}$. To perform the computation, we assume
that the perturbation of the phase of the asteroid is similar to a
Brownian motion $W\left(\frac{t}{\tau_{\varphi}}\right)$. However
we strongly emphasize that this particular ansatz for the perturbation
of the asteroid is chosen to perform analytic calculations, but while the exact result will depend on the particular expression
of $\varphi$, the order of magnitude of the result will not. Our result will thus give the correct order of magnitude of the diffusion coefficient as a function of the correlation time $\tau_{\varphi}$ of $\varphi$. 

The important result of the calculation of appendix \ref{sec:Computation-of-the}
is to show how the diffusion coefficient $D(\Lambda)$ scales with
$\tau_{\varphi}$. We have explained in Sect. \ref{sec:Diffusion-over-the}
that diffusion only occurs if the motion of the asteroid decorrelates
fast enough. It is therefore natural to expect that the diffusion
coefficient is larger when $\tau_{\varphi}$ is smaller. On the contrary,
if the motion of the asteroid is regular, there is no diffusion at
all, the diffusion coefficient should be zero for infinite $\tau_{\varphi}$.
A rough order of magnitude for the diffusion coefficient $D(\Lambda)$
from the formula (\ref{eq:diff}) of appendix \ref{sec:Computation-of-the}
writes in non dimensional variables
\begin{equation}
D(\Lambda)\propto\frac{\left|G\right|^{2}}{(n_p-\nu)^{2}\tau_{\varphi}},\label{eq:diff coef}
\end{equation}
where $\left|G\right|$ is the typical order of magnitude of the function
$G$, $n_p$ is the Keplerian pulsation of Mars, $\nu$ is the Keplerian pulsation of the asteroid, and
$\tau_{\varphi}$ is the Lyapunov time of the asteroid, all expressed in unit of time period of Mars. We can summarize this result by saying that the effect of the chaotic asteroid 
on the planet Mars is similar to a white noise process acting on the semi-major axis $a_p$ of Mars. In the  equation for $a_p$, the white noise term due to the asteroid writes
\begin{equation}
\frac{{\rm d} a_p}{{\rm d}t} =\sqrt{D}\xi(t),\label{eq:semi major}
\end{equation}
where $D$ can be written in dimensional units
\begin{equation}
D\propto 4 \left(\frac{m}{M_S}\right)^2\frac{\mathcal{G}M_S a_p}{|a_p-a|^2(n_p-\nu)^{2}\tau_{\varphi}}.\label{eq:dimension diff}
\end{equation}

Combining
expressions (\ref{eq:Diff_time}) and (\ref{eq:diff coef}) gives
the non-dimensional expression for the typical time  $\tau_{diff}$ after which Mars looses the memory of its initial longitude 
\begin{equation}
\tau_{diff}=\left(\frac{1}{3}\epsilon^{2}\left(\frac{\partial n_p}{\partial\Lambda}\right)^{2}\frac{\left|G\right|^{2}}{(n_p-\nu)^{2}\tau_{\varphi}}\right)^{-1/3}.\label{eq:tau_diff_nondimension}
\end{equation}
In order to derive this result, we have chosen the scaling of time, length and mass such that all
parameters are of order one, except the non-dimensional mass $\epsilon=\frac{m}{M_S}$
of the asteroid, and the non-dimensional Lyapunov time  of the asteroid $\tau_{\varphi}\simeq 10^4$ expressed in the units of time period of Mars.
To give the order of magnitude in non dimensional variables, expression (\ref{eq:tau_diff_nondimension}) can thus be simplified in
\begin{equation}
\tau_{diff}=\left(\frac{\epsilon^{2}}{\tau_{\varphi}}\right)^{-1/3},
\end{equation}
from which we deduce the dimensional expression of $\tau_{diff}$ in seconds
\begin{equation}
\tau_{\text{diff}}\propto\left(\left(\frac{m_{a}}{M_{S}}\right)^{2}\frac{\mathcal{G}M_{S}}{a_p^{3}\tau_{\varphi}}\right)^{-1/3}.
\end{equation}

\paragraph{Assumption of time scale separation}

Our model relies on two hypothesis. First, in order for the white noise limit in (\ref{eq:white noise})
to be valid, we have to assume that a time scale separation exists between the correlation time of
the asteroid and the diffusion time. Second, the semi-major axis should be constant to first
order in $\epsilon$ during the diffusion over the mean longitude.
We can summarize the time scale separation as the following inequalities
\begin{equation}
\tau_{\varphi}\ll\tau_{diff}\ll\tau_{a_p},\label{eq:time separation}
\end{equation}
where we have introduced $\tau_{a_p}$ a typical time of variation
of the semi-major axis.

To check the first inequality, we evaluate the ratio $\frac{\tau_{\varphi}}{\tau_{diff}}$
with (\ref{eq:tau_diff_dimension}). We find that $\frac{\tau_{\varphi}}{\tau_{diff}}\propto\left(\epsilon\frac{\tau_{\varphi}}{T_p}\right)^{\frac{2}{3}}$ where $T_p$ is the Keplerian period of Mars.
With the values of table (\ref{tab:Physical-properties-of}) $\epsilon=\frac{m}{M_S}\approx10^{-9}$
and $\frac{\tau_{\varphi}}{T_p}\approx10^{4}$, then the ratio is smaller than $10^{-3}$
and the first hypothesis is fulfilled. Eq. (\ref{eq:semi major}) together with (\ref{eq:dimension diff})
also shows that a typical diffusion time for the semi-major axis $a_p$ should be $\tau_{a_p}\propto\frac{a^2_p}{D}$. This gives the order of magnitude $\tau_{a_p}\approx 10^{22}$ yr and the ratio $\frac{\tau_{diff}}{\tau_{a_p}}$ is of the order of $\frac{\tau_{diff}}{\tau_{a_p}}\approx10^{-15}$. As
a consequence, we check that our simplified model (\ref{eq:simple Hamiltonian})
is consistent. 

The model does not conserve energy and angular momentum.
The conservation laws could be taken into account in principle by studying the back reaction of Mars on the asteroid at next order in the coupling between Mars and the asteroid. This would add a drift term in the stochastic equations (\ref{eq:stochastic system}) to balance
the long term variations of energy and angular momentum. However the mean values of energy and angular momentum are
expressed in terms of the semi-major axis $a_p$ which evolves on a timescale $\tau_{a_p}$ much longer than the timescale of interest ($\tau_{diff}$ ), that's why the integrals of motion can  be considered as constant. As a consequence the drift term is irrelevant on this timescale. We thus conclude that the time scale separation (\ref{eq:time separation}) ensures that our model has no bias because
of diffusion of energy and angular momentum, on the range of timescales of interest.

\paragraph{Action of several asteroids}

With the results of appendix \ref{sec:Computation-of-the}, we are
able to give the order of magnitude for the diffusion time of the mean
longitude of Mars. On timescales longer that their respective Lyapunov time, the motion of the asteroids can be considered as statistically independent. The perturbations of the asteroids on Mars can thus be considered separately. This allows to give an estimation of the total perturbation. 

\begin{table*}[!t]

\caption{Effect of the three largest asteroids on the mean longitude of Mars}
\label{tab:asteroids}

\begin{tabular}{c c c c c}
\hline 
 & nondimensional mass ($\epsilon$)$\times10^{-10}$ & Lyapunov time (yr) $\tau_{\varphi}$ & $\tau_{diff}$ (Myr) & extrinsic Lyapunov time $\tau_{e}$ (Myr)\\
\hline 
\hline 
Ceres & $4.7$ & 28900 & 22 & 25\\

Vesta & 1.3 & 14282 & 24 & 23\\

Pallas & 1.05 & 6283 & 33 & 36\\

total 3 asteroids & 7.05 & - & 17 & 18\\
\hline 
\end{tabular}

\end{table*}

Table (\ref{tab:asteroids}) is the first important quantitative result of this work. It gives
an estimation of the time we have to wait before seeing a noticeable
influence of the three largest asteroids on the mean longitude of Mars. The diffusion time
$\tau_{diff}$ depends strongly both on the mass of the asteroid and
on the Lyapunov time $\tau_\varphi$ of the asteroid. In particular we
see that Ceres and Vesta seem to have similar effects on Mars because
Ceres is much larger, but is also much less chaotic than Vesta.

\paragraph{Separation of the trajectories of Mars: super diffusion versus exponential separation}
We also report in the last column of table (\ref{tab:asteroids})
the extrinsic Lyapunov time $\tau_{e}$ of planets perturbed by the
asteroid belt. It is defined as the inverse of the largest Lyapunov
exponent of the dynamical system (\ref{eq:model}). Physically, it
corresponds to the typical time of exponential separation of two close
initial conditions in (\ref{eq:model}) for the mean longitude of Mars. We do not report details about the computation
of $\tau_{e}$ because the method is very similar to the computation of $\tau_{diff}$ and is already explained in the work of \cite{lam2014stochastic}.

The Lyapunov time $\tau_{e}$ should be compared to $\tau_{diff}$ associated to the superdiffusion of
the mean longitude of the planet. Those two times are the consequence of two different mechanisms: As we have explained
in Sect. \ref{sec:A-simplified-model}, the superdiffusion occurs as a consequence of the chaotic dynamics of the asteroids, when averaging over a large number of initial
phases of the asteroid, whereas $\tau_{e}$  is the time which characterizes the exponential separation between two trajectories with slightly different initial
planetary longitudes, with \emph{one single} realization of the asteroid perturbation.
While those two times are the consequence of two different mechanisms, the reader
can notice on table (\ref{tab:asteroids}) that they surprisingly have the same order of magnitude.

\begin{figure}[h]
\includegraphics[scale=0.5]{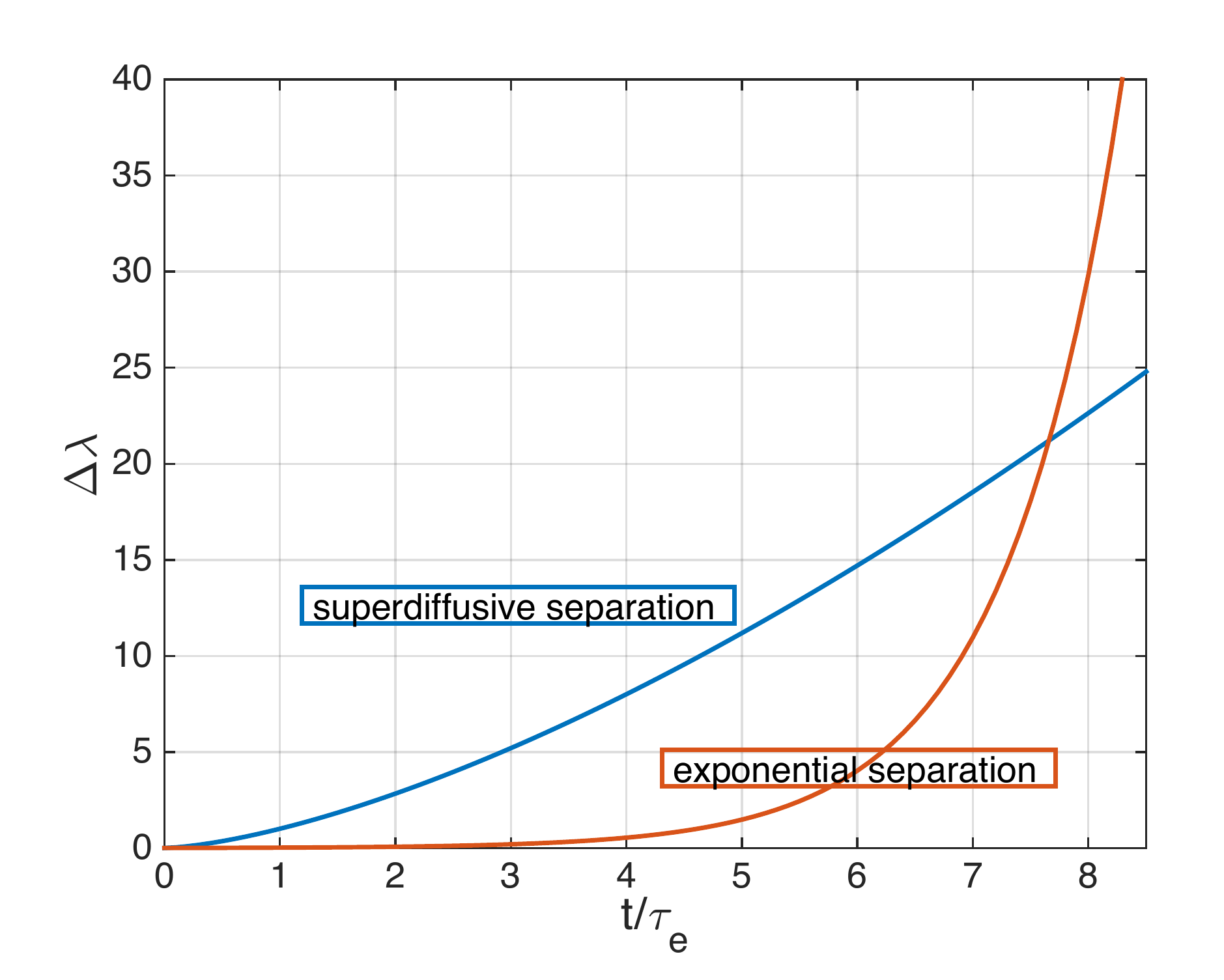}
\caption{The separation between the trajectories of Mars with close initial
conditions can occur with two mechanisms. The superdiffusion mechanism
represented by the blue curve gives a separation scaling as a power-law
$\left(\frac{t}{\tau_{diff}}\right)^{3/2}$. The chaos in the system
separates two trajectories exponentially fast as displayed by the
red curve. Here we have chosen an initial separation $\Delta\lambda_{0}\approx0.1$.
The superdiffusion mechanism is the main one for short times. \label{fig:separation mechanisms}}
\end{figure}

Although the diffusion time $\tau_{diff}$ and the extrinsic Lyapunov time $\tau_{e}$ are
of same order, the superdiffusion is the main contribution to the
separation of trajectories. Indeed, if one considers two trajectories
for the planet Mars initially separated by a difference in longitude
$\Delta\lambda_{0}\ll1$, the superdiffusion mechanism amplifies the
initial separation after a time $t$ as
\begin{equation}
\Delta\lambda(t)\sim\Delta\lambda_{0}+\left(\frac{t}{\tau_{diff}}\right)^{3/2}.\label{eq:polynomial separation}
\end{equation}
On the other hand, the exponential separation because of the chaos
created by the asteroid writes 
\begin{equation}
\Delta\lambda(t)\sim\Delta\lambda_{0}\exp\left(\frac{t}{\tau_{e}}\right).\label{eq:exponential separation}
\end{equation}
Remember that $\tau_{e}\approx\tau_{diff}$. As shown on Fig. (\ref{fig:separation mechanisms}), for times smaller or of the same order of $\tau_{e}$, the power-law growth  overcomes the exponential growth. That's why the superdiffusion mechanism is more relevant than exponential
separation for the computation of the probability distribution of planetary mean longitudes. 

The first conclusion we can give at this stage is that the characteristic time scale over which asteroids affect the planet longitudes is of order of $10$ Myr. In order to find this order of magnitude, we have made a computation with  circular and coplanar orbits (see Sect.~\ref{sec:A-simplified-model}). However we stress that taking into account of the eccentricity or the inclination of the planetary orbit would not change this order of magnitude. Moreover, taking into account the secular motion of the eccentricities and inclinations would neither change this order of magnitude. Indeed the mechanism that causes longitude diffusion is the perturbation by the asteroids of the planetary semi-major axis, while on secular time scales the semi-major axis is an adiabatic invariant and does not evolve due to the secular evolution of the planets. 

This superdiffusion timescale for the planetary longitude, of order of $10$ Myr, is of the same order of magnitude as the Lyapunov time $\tau_{i}$ for the orbital parameters of the inner solar system, as given by \cite{Laskar1989}. It is thus a natural question to study the stochastic effects of the asteroids on the secular orbital parameters themselves. This is the subject of next section. 

\section{Asteroid influence on secular motions\label{sec:Influence-on-secular}}

In Sect. \ref{sec:Orders-of-magnitude}, we have explained the effect of
the gravitational influence of asteroids on planetary mean longitudes is a superdiffusion, which means
that the standard deviation of the mean longitude grows with time as $\left(\frac{t}{\tau_{diff}}\right)^{\frac{3}{2}}$.  However an incertitude on the exact position of the planet
on its orbit does not change the secular equations, and the evolution of the orbital parameters  inclinations $i$ and
eccentricities $e$ of planetary orbits. The change of the orbital parameters are the most important properties that affect many applications, for instance climate. We are thus naturally led to consider the influence of asteroids on the orbital parameters.

The secular equations with the contribution of asteroids can be described
very formally by a Hamiltonian composed of three terms
\[
H_{sec}=H_{LL}+H_{NL}+H_{a},
\]
where $H_{LL}$ is the Hamiltonian of Laplace\textendash Lagrange with all quadratic
terms in $e$ and $i$, $H_{NL}$ gathers the terms of order larger
than $3$ in $e$ and $i$ and $H_{a}$ describes the interaction with the
asteroids. We are exactly in the framework described in Sect. \ref{sec:The-theoretical-framework}: 
the secular Hamiltonian $H_{LL}+H_{NL}$ without asteroids is chaotic with an intrinsic Lyapunov time $\tau_i$.
The Hamiltonian $H_{LL}+H_{a}$ describes the integrable motion of Laplace\textendash Lagrange perturbed by the gravitational interaction with asteroids.
The perturbation by asteroids $H_{a}$ is an external source  of chaos in the integrable system, and causes exponential separation of trajectories with close initial conditions over a time $\tau_e$. We want to know which time is larger between $\tau_i$ and $\tau _e$. This amounts to determine if the external source of chaos dominates compared to the intrinsic chaos of the secular system. The other related question is to know wether a superdiffusion mechanism occurs on the secular orbital parameters, similar to the one that happens for planetary mean longitudes, and on which timescale $\tau_{sec}$ the superdiffusion mechanism perturbs  the secular motion.

As we have done in the last sections, we will focus only on orders of magnitude. The present section
is again somehow technical, and is mostly an extension of the techniques
already developed in Sect. \ref{sec:Diffusion-over-the}. The reader
can skip it and go directly to the conclusions presented in the next
section.

Let $\alpha$ be the set of all canonical variables $e\exp(j\varpi)$, $\sin\left(\frac{i}{2}\right)\exp(j\Omega)$ for all planets of the solar system, and their complex conjugates, where $j=\sqrt{-1}$, $\varpi$ is the longitude of the perihelion, and $\Omega$ is the longitude of the ascending node (see e.g. \cite{murray1999solar} p.48). $\alpha$ is a vector of dimension 32, taking into account the 8 planets. The Laplace\textendash Lagrange system
writes
\begin{equation}
\dot{\alpha}=A(\Lambda)\alpha\label{eq:LaplaceLagrange}
\end{equation}
where $\Lambda$ is now the $8$ dimensional vector of the canonical
momentums associated to planetary mean longitudes. In the secular equations,
$\Lambda$ is a constant because the mean longitudes $\lambda$ do
not appear any more in the Hamiltonian after the averaging procedure.
The gravitational interaction with asteroids is described by the Hamiltonian
$h_a:=\underset{p}{\sum}\frac{-\mathcal{G}m_{p}m}{\left|\mathbf{r}_{p}-\mathbf{r}\right|}$,
and $H_{a}$ is the secular contribution of this Hamiltonian. The
perturbation induced by asteroids perturbs the dynamics (\ref{eq:LaplaceLagrange}) by two different mechanisms:
\begin{enumerate}
\item The asteroid motion gives a white noise term in the equation on the semi-major axis (see (\ref{eq:semi major})), and thus on 
$\Lambda_p:=m_p\sqrt{\mathcal{G}M_Sa_p}$ as was explained in Sect. \ref{sec:Diffusion-over-the}.
The matrix $A$ of the secular system (\ref{eq:LaplaceLagrange})
should not be considered as a constant in the secular equations, because
we have to take into account the diffusion over $\Lambda$. We show
in the following that the vector $\alpha$ of orbital parameters is
stochastic and follows a \emph{multidimensional }superdiffusion, through
a mechanism very similar to the superdiffusion of the mean longitudes.
\item The secular Hamiltonian $H_{a}$ creates a \emph{direct} additional term in the equation. It creates terms involving the derivatives of $H_a$ in the secular equations. In principle, those terms could be analyzed by the method developed is Sect.~\ref{sec:Diffusion-over-the}. If we write $H_a$ in action-angle variables, we can isolate the average contribution (coming from an averaging procedure over the secular angles), and the diffusive contribution, beyond the averaging principle. The computation of the diffusion coefficient involves the correlation time of the orbital parameters of asteroid's orbits. We do not know precisely the order of magnitude of this time, that's why it seems difficult to give the order of magnitude of this diffusion coefficient. We assume that this direct diffusion mechanism is smaller than the superdiffusion mechanism coming from diffusion of the planetary semi-major axes.
\end{enumerate}

Let $A(t)=A(\Lambda(t))$ and $u(t)=e^{-A_{0}t}\alpha(t)$, where
$A_{0}:=A(0)$. We thus have $u(0)=\alpha(0)$. With this change of
variable, we integrate out the fast motion coming from $A_{0}$. $u(t)$
satisfies the differential equation
\[
\dot{u}(t)=e^{-A_{0}t}\delta A(t)e^{A_{0}t}u(t)
\]
where we have set $\delta A(t)=A(t)-A_{0}$. We expect $\delta A(t)$
to be small because it comes from the variation of $\Lambda$. We remember that $\epsilon=m/M_s$  quantify the asteroid effect on the planets. When
$\epsilon$ goes to zero, we thus have $\delta A\rightarrow0$. This
allows us to do a computation to order 1 in $\epsilon$ and we have
\begin{equation}
u(t)\simeq\alpha(0)+\int_{0}^{t}{\rm d}s~e^{-A_{0}s}\delta A(s)e^{A_{0}s}\alpha(0).\label{eq:secular diff}
\end{equation}

The difference $\delta u:=u(t)-\alpha(0)$ comes from the diffusive
behavior of the set of variables $\alpha(t)$. $\delta A$ is small,
and the first order is given by a linear expansion of the matrix $A$
w.r.t $\Lambda$. We have $\delta A=\frac{\partial A}{\partial\Lambda}\delta\Lambda$.
According to Sect. \ref{sec:Diffusion-over-the}, $\delta\Lambda(t)$
is a N-dimensional Brownian motion - N being the number of planets
- with magnitude $\epsilon\sqrt{D(\Lambda)}$ where $D$ is now a
diagonal matrix gathering the coefficients $D_{p}(\Lambda_{p})$ for
each planet $p$. We take for $\delta\Lambda$ the expression $\delta\Lambda=\epsilon\sqrt{D(\Lambda)}W(t)$, which is
valid only on the Myr time scale. Finally, we estimate the diffusion
time of the set $\alpha$ by computing explicitly the quantity $\mathbb{E}\left[\delta u^{\dagger}\delta u\right]$
($\dagger$ stands for the complex conjugated and transposed of a
matrix or a vector). The computation is done in appendix \ref{sec:Diffusion-of-secular}.
We show that $\mathbb{E}\left[\delta u^{\dagger}\delta u\right]$
has a component growing like $t^{3}$ which is the signature of a
superdiffusive behavior. 

The time scale for the superdiffusion of orbital elements $e,i,\varpi,\Omega$ writes in non-dimensional units
\begin{equation}
\tau_{sec}\simeq\left(\frac{1}{3}\epsilon^{2}\underset{l,p}{\sum}D_{p}\left|\alpha_{l}^{\dagger}\frac{\partial A_{0}}{\partial\Lambda_{p}}\alpha_{l}\right|^{2}\right)^{-\frac{1}{3}},\label{eq:tau_sec_nondim}
\end{equation}
where the $\alpha_{l}$ are the eigenvectors of the matrix of secular
frequencies $A_{0}$. As one can see, the mechanism of superdiffusion
of orbital elements is similar to the superdiffusion  of the mean longitudes.

Comparing  expression (\ref{eq:Diff_time}) for $\tau_{diff}$ and expression (\ref{eq:tau_sec_nondim}) for $\tau_{sec}$, one sees that the
nonisochronic parameter $\frac{\partial n_p}{\partial\Lambda}$  is replaced here by the elements
of the derivative $\frac{\partial A_{0}}{\partial\Lambda}$. The latter
matrix is essentially the matrix of the derivatives of secular frequencies. Therefore, the mechanism of superdiffusion for the secular orbital parameters is similar to the one for the mean longitudes, because in both cases it comes from a diffusion of the eigenfrequencies of the integrable motion. $\tau_{diff}$ comes from the diffusion of the Keplerian pulsation $n_p$, and $\tau_{sec}$ comes from the diffusion of the eigenfrequencies $g$ and $f$.

To evaluate an order of magnitude of $\tau_{sec}$, we do the rough
assumption that the coefficients $\alpha_{l}^{\dagger}\frac{\partial A_{0}}{\partial\Lambda_{p}}\alpha_{l}$
are essentially of the order of magnitude of the secular frequencies, i.e of few arcseconds/yr.
This is the major difference with the superdiffusion of mean longitudes.
In expression (\ref{eq:dimension diff}), the Keplerian frequency is of
order of the arc/yr. Because of the difference between the frequencies
of Keplerian motion and the secular eigenfrequencies
$g$ and $f$ for the solar system, the time scale $\tau_{sec}$ for
a superdiffusion of the orbital parameters is expected to be much
larger than the diffusion time of the mean longitudes. With dimensional
variables, the order of magnitude of the time $\tau_{sec}$ is expressed
as
\begin{equation}
\tau_{\text{sec}}\propto\left(\left(\frac{m_{a}}{M_{S}}\right)^{2}\frac{\nu_{sec}^{2}}{\tau_{\varphi}}\right)^{-1/3},
\end{equation}
and we find that $\tau_{sec}$ should
be typically larger than 10 Gyr. This time is also the order of magnitude of the Lyapunov time $\tau_e$  associated with the chaos created by
asteroids on the secular system of Laplace\textendash Lagrange.  

We thus conclude that $\tau_e$  is much larger than $\tau_i$, the Lyapunov time for the eccentricities and inclinations of the internal planets. The asteroids provide thus a very small perturbation to the intrinsically chaotic secular dynamics. 

\section{Conclusions\label{sec:Conclusion}}

In the present paper, we have investigated the influence of asteroids on
the long-term dynamics of the solar system. With the technique of
stochastic averaging, we have shown that the chaotic
behavior of the main asteroids in the asteroid
belt: Ceres, Vesta and Pallas, perturbs the long term dynamics of planets, and give a stochastic
contribution to the motion of planets. We have explained that the main
effect is a superdiffusion of planet mean longitudes, i.e
a variance of the mean longitudes growing like $\left(\frac{t}{\tau_{diff}}\right)^{3}$
instead of $\frac{t}{\tau_{diff}}$ for normal diffusion. An order of magnitude of $\tau_{diff}$ is given by (\ref{eq:tau_diff_dimension}). We have found that
for the planet Mars perturbed by the asteroid belt, the superdiffusion
becomes significative after 10 Myr. This order of magnitude should
be the same for all planets in the inner solar system. This result thus confirms on a theoretical ground that no accurate planetary ephemerides can be elaborated for times larger that $10$ Myr, as was already observed in the numerical simulations of  \cite{laskar_Gastineau_2011_AA_strong}.

Within the Myr time scale, the motion over the orbit is averaged
out and the dynamics is described by the secular equations. We have studied
the effect of asteroids on secular motions and we have found that the mechanism
of superdiffusion also happens on the planet eccentricities
and inclinations. The time $\tau_{sec}$ that characterizes this superdiffusion, given by (\ref{eq:tau_sec_dimension}), has been estimated to 10 Gyr. The intrinsic chaotic dynamics of the secular equations occurs within a time $\tau_{i}$ of order of 10 Myr. As a consequence the
evolution of the planetary eccentricities and inclinations probability distributions, over
one Gyr time scale, is completely dominated by the intrinsic secular chaos. At this stage, the conclusion is that asteroids have an
influence on secular motion so small that it should change the
statistical distributions of eccentricities and inclinations  computed with the secular
equations by \cite{laskar2008chaotic}, only through a tiny perturbation. 

\begin{acknowledgements}
We thank J. Laskar for his helpful advises during the preliminary
stage of this work. The research leading to these results has received
funding from the European Research Council under the European Union's
seventh Framework Program (FP7/2007-2013 Grant Agreement No. 616811).
 
\end{acknowledgements}

\bibliographystyle{aa} 
\bibliography{biblio_celeste.bib} 

\begin{appendix}

\section{Computation of the diffusion coefficient\label{sec:Computation-of-the}}

We want to give an order of magnitude of the Lyapunov time (\ref{eq:Diff_time})
of the longitude $\lambda$ of Mars feeling the perturbation of a
large asteroid of the asteroid belt. The perturbative function writes
\[
G(\Lambda,\lambda-\varphi)=\frac{-1}{\left|\Lambda^{2}e^{i(\lambda-\varphi)}-a\right|}
\]
where $\Lambda$ is the conjugated momentum of the mean longitude
$\lambda$. In the following, we will always assume that $\Lambda^{2}<a$.
The Fourier expansion of $G$ has been given in the general case by
\cite{laskar2010explicit}, and writes in our simplified model
\begin{eqnarray}
\widehat{G}(k) & = & \frac{1}{\Lambda_{0}^{2}}\stackrel[q=k]{N}{\sum}\left(\frac{\Lambda_{M}}{\Lambda_{0}}\right)^{4q-2k}f_{2q-k,q}\label{eq:Fourier coef}\\
f_{n,q} & = & \frac{(2q)!(2n-2q)!}{2^{2n}(q!)^{2}[(n-q)!]^{2}}\nonumber 
\end{eqnarray}
where $N$ is a cut off to stop the Fourier expansion. In our calculations,
we took $N=20$.

We have the Fourier decomposition of $G$ in the form $G=\underset{k}{\sum}\widehat{G}(k)e^{ik(\lambda-\varphi)}$.
Our aim is to evaluate the quantity $\mathbb{E}\left[\frac{\partial G}{\partial\lambda}(t)\frac{\partial G}{\partial\lambda}(0)\right]$
where for simplicity we use the shortcoming $G(t)=G(\Lambda,\lambda(t)-\varphi(t))$.
The reader should bear in mind that $\lambda(t)$ is given by the
\emph{unperturbed }dynamics, because to compute the fast motions we
have to ``freeze'' the slow variables. If we freeze $\Lambda$ in
the dynamics of $\lambda$, we simply get the Keplerian motion. Therefore
$\lambda(t)=\lambda_{0}+nt$. The chaotic dynamics of the asteroid
is modeled by $\varphi(t)=\varphi_{0}+\nu_{a}t+\delta\varphi(t)$
where $\delta\varphi$ should account for the chaotic diffusion on
a time scale $\tau_{\varphi}$. We will give its expression later
on. We thus have
{\small 
\begin{eqnarray*}
\mathbb{E}\left[\frac{\partial G}{\partial\lambda}(t)\frac{\partial G}{\partial\lambda}(0)\right] & = & \underset{k,k'}{\sum}kk'\widehat{G}(k)\widehat{G}(k')^{*}\mathbb{E}\left[e^{i(k\lambda(t)-k'\lambda(0))}e^{-i(k\varphi(t)-k'\varphi(0))}\right]\\
 & = & \underset{k,k'}{\sum}kk'\widehat{G}(k)\widehat{G}(k')^{*}e^{ikt(n-\nu_{a})}...\\
 &&...\mathbb{E}\left[e^{i(\lambda_{0}-\varphi_{0})(k-k')}\right]\mathbb{E}\left[e^{-ik\delta\varphi(t)}\right].
\end{eqnarray*}}
We have two averages to perform. The invariant measure for the initial
conditions $\lambda_{0}$ and $\varphi_{0}$ is the uniform measure
over the range $[0,2\pi]$. Thus the term $\mathbb{E}\left[e^{i(\lambda_{0}-\varphi_{0})(k-k')}\right]=\delta_{k-k'}$.
What is then the Esperance of $e^{-ik\delta\varphi}$ ? For a general
chaotic trajectory, it should be a complicated function, decreasing
with the time scale $\tau_{\varphi}$. We chose for $\delta\varphi$
the Brownian motion $W\left(\frac{t}{\tau_{\varphi}}\right)$ which
has a Gaussian statistics. It comes 
\begin{eqnarray*}
\mathbb{E}\left[e^{-ik\delta\varphi(t)}\right] & = & \frac{1}{\sqrt{2\pi t/\tau_{\varphi}}}\int{\rm d}xe^{-ikx}e^{-\frac{x^{2}}{2t/\tau_{\varphi}}}\\
 & = & e^{-\frac{t}{2\tau_{\varphi}}k^{2}}.
\end{eqnarray*}
Finally, the expression of our correlation function is 
\[
\mathbb{E}\left[\frac{\partial G}{\partial\lambda}(t)\frac{\partial G}{\partial\lambda}(0)\right]=\underset{k}{\sum}k^{2}\left|\widehat{G}(k)\right|^{2}e^{-\frac{t}{2\tau_{\varphi}}k^{2}}e^{ik(n-\nu_{a})t}.
\]
It remains to integrate this expression over time according to (\ref{eq:coefficients}),
and we obtain
\begin{equation}
D(\Lambda)=\underset{k}{\sum}k^{2}\left|\widehat{G}(k)\right|^{2}\frac{1}{(\nu_{a}-n)^{2}\tau_{\varphi}}\frac{1}{\left(\frac{k}{2(\nu_{a}-n)\tau_{\varphi}}\right)^{2}+1}.\label{eq:diff}
\end{equation}
Expressions (\ref{eq:Fourier coef}) and (\ref{eq:diff}) allow to
compute the magnitude of the diffusion coefficient and then to estimate
the diffusion time scale of the mean longitude of Mars. 

\section{Diffusion of secular variables\label{sec:Diffusion-of-secular}}

Consider again equation (\ref{eq:secular diff}) which writes
\[
\delta u(t)\simeq\epsilon\int_{0}^{t}{\rm d}s~e^{-A_{0}s}\frac{\partial A_{0}}{\partial\Lambda_{p}}(s)\sqrt{D(\Lambda_{p})}W_{p}(s)e^{A_{0}s}u_{0}
\]
with a summation on the index $p$. And we want to compute
{\small
\begin{eqnarray*}
\mathbb{E}\left[\delta u^{\dagger}\delta u\right]=\epsilon^{2}\iint_{0}^{t}{\rm d}s{\rm d}s'\mathbb{E}\left[W_{p}(s)W_{p'}(s')\right]\sqrt{D\left(\Lambda_{p}\right)D\left(\Lambda_{p'}\right)}...\\
...u_{0}^{\dagger}e^{A_{0}^{\dagger}s'}\frac{\partial A_{0}^{\dagger}}{\partial\Lambda_{p}}e^{-A_{0}^{\dagger}s'}e^{-A_{0}s}\frac{\partial A_{0}}{\partial\Lambda_{p'}}e^{A_{0}s}u_{0}.
\end{eqnarray*}}
Then we use the fact that $A_{0}$ is an anti-hermician operator,
and therefore $A_{0}^{\dagger}=-A_{0}$. The Brownian motions $W_{p}$
are independent, thus we have $\mathbb{E}\left[W_{p}(s)W_{p'}(s')\right]=\delta(p-p')\inf(s,s')$.
We get
{\small
\begin{eqnarray*}
\mathbb{E}\left[\delta u^{\dagger}\delta u\right]=\epsilon^{2}\iint_{0}^{t}{\rm d}s{\rm d}s'\inf(s,s')D\left(\Lambda_{p}\right)u_{0}^{\dagger}e^{-A_{0}s'}\frac{\partial A_{0}^{\dagger}}{\partial\Lambda_{p}}e^{-A_{0}(s-s')}\frac{\partial A_{0}}{\partial\Lambda_{p}}e^{A_{0}s}u_{0}.
\end{eqnarray*}}
Be careful that in the preceding expression, $A_{0}$ do not commute
with its derivative $\frac{\partial A_{0}}{\partial\Lambda}$. Let
$\alpha_{i}$ be a set of eigenvectors of $A_{0}$ with eigenfrequencies
$j\nu_{i}$, such that $e^{A_{0}s}\alpha_{i}=e^{j\nu_{i}s}\alpha_{i}$.
We decompose $u_{0}$ on this set of eigenvectors, $u_{0}=\sum c_{i}\alpha_{i}$.
Finally, we introduce the closure relation $Id=\sum\alpha_{i}\alpha_{i}^{\dagger}$.
We have
{\small 
\begin{eqnarray*}
\mathbb{E}\left[\delta u^{\dagger}\delta u\right]=\epsilon^{2}\underset{i,k,l}{\sum}c_{k}^{*}c_{l}\int_{0}^{t}\int_{0}^{t}{\rm d}s{\rm d}s'\inf(s,s')D\left(\Lambda_{p}\right)\alpha_{k}^{\dagger}e^{-j\nu_{k}s'}...\\
...\frac{\partial A_{0}^{\dagger}}{\partial\Lambda_{p}}\alpha_{i}e^{-j\nu_{i}(s-s')}\alpha_{i}^{\dagger}\frac{\partial A_{0}}{\partial\Lambda_{p}}e^{j\nu_{l}s}\alpha_{l}.
\end{eqnarray*}}
We have to evaluate a double integral of the form $\iint{\rm d}s{\rm d}s'\inf(s,s')e^{i(\nu_{l}-\nu_{i})s}e^{-i(\nu_{k}-\nu_{i})s'}$,
which does not present any difficulty. The result is that this double
integral grows like $t$ unless $i=k=l$ and in that case the growth
scales like $t^{3}$. This means that 
\[
\mathbb{E}\left[\delta u^{\dagger}\delta u\right]\underset{\nu t\gg1}{\sim}\frac{1}{3}\epsilon^{2}\underset{i}{\sum}\left|c_{i}\right|^{2}D\left(\Lambda_{p}\right)\left|\alpha_{i}^{\dagger}\frac{\partial A_{0}}{\partial\Lambda_{p}}\alpha_{i}\right|^{2}t^{3},
\]
 this asymptotic behavior is valid for times $t$ large in front of
$\frac{1}{\nu_{sec}}$ where $\nu_{sec}$ is the typical order of magnitude of the frequencies of the
Laplace\textendash Lagrange system, and $t$ is small compared to $\frac{1}{\epsilon^{2}\left\Vert D\left(\Lambda\right)\right\Vert }$.
The latter assumption is easily satisfied because $\epsilon<10^{-9}$
and the former assumption is satisfied for times larger than a Myr.

\section{Notations\label{sec:Notations}}
\begin{table*}
\begin{tabular}{c c}
\hline
$M_S$ & mass of the sun\\
$m_p$ & mass of Mars\\
$m$ & mass of the asteroid\\
$a_p$ & semi-major axis of Mars\\
$a$ & semi-major axis of the asteroid\\
$\mathcal{G}$ & gravitational constant\\
$G$ & gravitational potential between the asteroid and Mars\\
$n_p$ & Keplerian pulsation of Mars\\
$T_p$ & Keplerian period of Mars\\
$\nu$ & Keplerian pulsation of the asteroid\\
$\lambda$ & mean longitude of Mars\\
$\varphi$ & mean longitude of the asteroid\\
$\Lambda$ & canonical momentum conjugated to $\lambda$\\
$\epsilon=m/M_S$ & reduced mass of the asteroid\\
$\tau_{\varphi}$ & Lyapunov time of the asteroid\\
$\tau_{diff}$ & typical time of superdiffusion of the mean longitude of Mars\\
$\tau_i$ & intrinsic Lyapunov time of the solar system\\
$\tau_e$ &  Lyapunov time of the planet Mars perturbed by the chaotic asteroid\\
$\nu_{sec}$ & typical frequency of the Laplace\textendash Lagrange system\\
$\alpha$ & vector of canonical variables of the Laplace\textendash Lagrange system\\
$A$ & matrix of the Laplace\textendash Lagrange system\\
$\tau_{sec}$ & typical time of superdiffusion of planetary orbital elements $e,i,\varpi,\Omega$\\ 
\hline
\end{tabular}
\end{table*}
\end{appendix}

\end{document}